\documentclass[twocolumn,aps,prb,superscriptaddress,floatfix,]{revtex4-1}
\usepackage{graphicx}
\usepackage{color}
\usepackage{ragged2e}
\usepackage[normalem]{ulem}
\usepackage{float}

\def\LSCO{La$_{5/3}$Sr$_{1/3}$CoO$_4$}
\def\LSCOx{La$_{2-x}$Sr$_{x}$CoO$_4$}
\def\LSNOx{La$_{2-x}$Sr$_{x}$NiO$_4$}

\begin{document}

\title{Direct evidence for charge stripes in a layered cobalt oxide}


%
%
%
%
%
%
%
%
%

\author{P. Babkevich}
\email{peter.babkevich@epfl.ch}
\affiliation{Laboratory for Quantum Magnetism, Institute of Physics, \'{E}cole Polytechnique F\'{e}d\'{e}rale de Lausanne (EPFL), CH-1015 Lausanne, Switzerland}
\author{P. G. Freeman}
\affiliation{Jeremiah Horrocks Institute for Mathematics, Physics and Astronomy,University of Central Lancashire, Preston, Lancashire, PR1 2HE, UK}
\author{M. Enderle}
\affiliation{Institut Laue--Langevin, CS 20156, 38042 Grenoble Cedex 9, France}
\author{D. Prabhakaran}
\affiliation{Department of Physics, University of Oxford,Clarendon Laboratory, Oxford, OX1 3PU, UK}
\author{A. T. Boothroyd}
\email{a.boothroyd@physics.ox.ac.uk}
\affiliation{Department of Physics, University of Oxford,Clarendon Laboratory, Oxford, OX1 3PU, UK}

\begin{abstract}
{\bf
Recent experiments indicate that static stripe-like charge order is generic to the hole-doped copper oxide superconductors and competes with superconductivity. Here we show that a similar type of charge order is present in \LSCO, an insulating analogue of the copper oxide superconductors containing cobalt in place of copper.  The stripe phase we have detected is accompanied by short-range, quasi-one-dimensional, antiferromagnetic order, and provides a natural explanation for the distinctive hourglass shape of the magnetic spectrum previously observed in neutron scattering measurements of \LSCOx\ and many hole-doped copper oxide superconductors. The results establish a solid empirical basis for theories of the hourglass spectrum built on short-range, quasi-static, stripe correlations.
}
\end{abstract}

\date{\today}
\maketitle


The hourglass spectrum has been observed in measurements on \LSCOx\ with $x=0.25-0.4$ (refs \onlinecite{boothroyd-nature-2011,gaw-prb-2013,drees-nature-2013}), but its origin has become a matter of contention. Initially\cite{boothroyd-nature-2011}, it was explained by a model of short-range spin and charge stripe correlations similar to those found in certain hole-doped copper-oxide superconductors near one-eighth doping\cite{tranquada-nature-1995}. However, the subsequent failure to observe charge stripe order in diffraction experiments led Drees {\it et al.} to propose a stripe-free model based on a nanoscopically phase-separated ground state comprising coexisting regions with local composition $x=0$ and $x=0.5$ (ref.~\onlinecite{drees-nature-2013}). The presence or absence of charge stripes, therefore, is of crucial importance for understanding the hourglass spectrum in \LSCOx.

The crystal structure of \LSCOx\ contains well-separated CoO$_2$ layers with  Co arranged on a square lattice. These layers are isostructural with the CuO$_2$ layers in the copper oxide high-temperature superconductors. The parent compound La$_2$CoO$_4$ is an insulator with N\'{e}el-type antiferromagnetic (AFM) order below $T_{\rm N} = $275\,K (ref.~\onlinecite{yamada-prb-1989}). Substitution of Sr for La adds positive charge (holes) onto the CoO$_2$ layers, converting Co$^{2+}$ into Co$^{3+}$ and suppressing the AFM order.
In the compositions of interest here the Co$^{2+}$ ions have spin $S=3/2$ and the Co$^{3+}$ ions carry no spin ($S=0$) \cite{hollmann-newjourn-2008,boothroyd-nature-2011}.

\begin{figure}
\includegraphics[width=0.9\columnwidth]{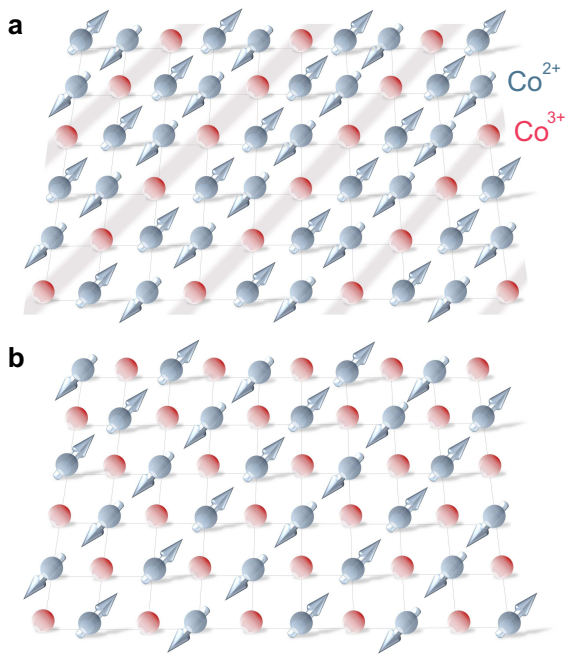}
\caption{\textbf{Models of ideal spin and charge order in \LSCOx\ } ({\bf a}) Period-3 spin and charge stripe order for $x = 1/3$. ({\bf b}) Checkerboard spin and charge order for $x = 1/2$. The arrows represent magnetic moments on Co$^{2+}$ ions and the red spheres denote Co$^{3+}$. The ions are located at the corners of the square lattice motif.  The same patterns of spin and charge order apply to \LSNOx. }
\label{fig1}
\end{figure}

\LSCOx\ compounds with $x \gtrsim 0.25$ exhibit diagonally-modulated incommensurate AFM order with a modulation period which varies linearly with doping from $x \simeq 0.25$ up to at least $x = 0.5$ (refs~\onlinecite{cwik-prl-2009,gaw-prb-2013}). Such magnetic behaviour is also found in the isostructural layered nickelates \LSNOx, where it is caused by ordering of the holes into stripes which align at 45$^{\circ}$ to the Ni-O bonds and which act as antiphase domain walls in the AFM order  \cite{chen-prl-1993,tranquada-prl-1994,yoshizawa-prb-2000,ulbrich-physicaC-2012}. The superstructure of spin and charge order observed in the nickelate composition of most relevance to the present work ($x=1/3$) has three times the period of the square lattice. For  \LSNOx\ with $x > 1/3$, the charge stripe period decreases linearly with increasing $x$ until at $x=0.5$ the charge order approaches an ideal checkerboard pattern, see Fig.~\ref{fig1}.

The linear doping dependence of the incommensurate magnetism observed in \LSCOx\ for $x<1/2$ was interpreted\cite{cwik-prl-2009} as possible evidence for the existence of diagonal spin-charge stripe phases like those found in \LSNOx. Consistent with this, indirect evidence for charge freezing at $\sim 100$\,K was inferred from magnetic resonance experiments on \LSCOx\  (refs~\onlinecite{lancaster-prb-2014,williams-arxiv-2015}). Until now, however, attempts to detect charge stripe order in the cobaltates more directly by neutron and X-ray diffraction have been unsuccessful. In fact, instead of the stripe phase expected by analogy with \LSNOx, recent diffraction measurements\cite{drees-nature-2014} made on \LSCO\ detected the presence of short-range checkerboard charge order, a very stable phase previously found in \LSCOx\ at $x=1/2$ (refs~\onlinecite{zaliznyak-prl-2000,zaliznyak-prb-2001}).

In this work we used polarized neutron diffraction to detect spin and charge stripes in \LSCO. The stripes have the same arrangement of spin and charge order (SO and CO) as those found in La$_{2-x}$Sr$_x$NiO$_4$ with $x \simeq 1/3$ but have a higher degree of disorder. The presence of this stripe phase provides a natural explanation for the hourglass magnetic spectrum of \LSCO.

\section*{Results}
\subsection*{Signatures of checkerboard and stripe correlations}

Neutrons do not couple directly to charge but can probe CO through the associated structural distortions. Polarization analysis was employed in this work to achieve an unambiguous separation of the structural and magnetic scattering from \LSCO\ due to charge and spin order (see Methods). The CO and SO phases of most relevance to our diffraction results are depicted in Fig.~\ref{fig1}. These are the period-3 stripe phase and the checkerboard phase.

\begin{figure*}
\includegraphics[width=0.7\textwidth]{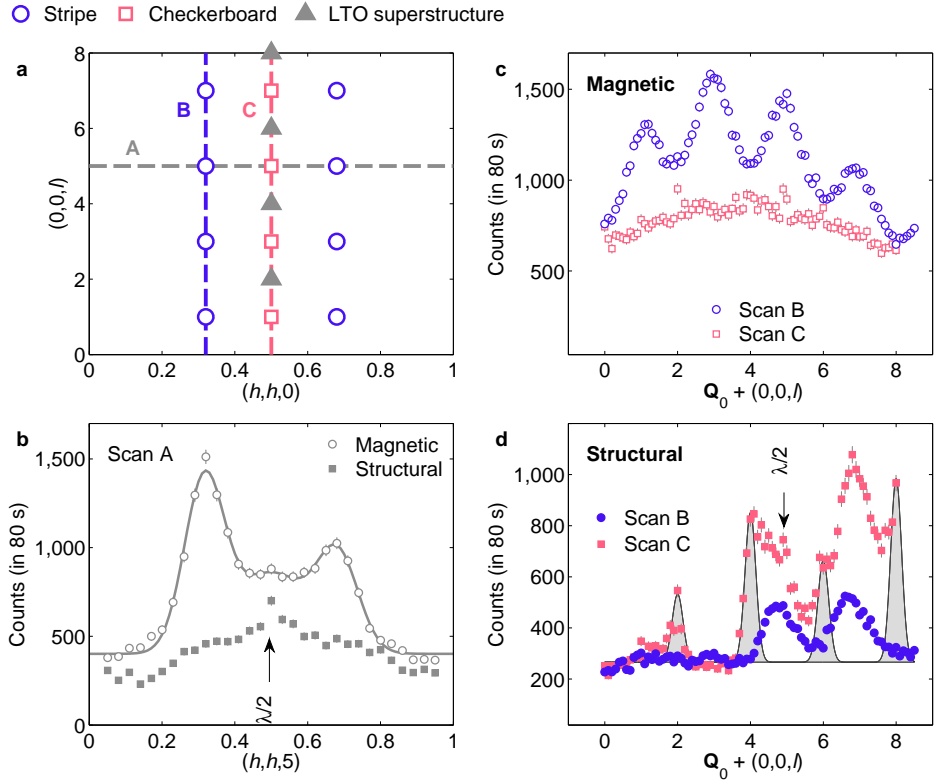}
\caption{\textbf{Neutron diffraction from \LSCO\ at 2\,K} ({\bf a}) Diagram of the $(h,h,l)$ reciprocal lattice plane showing positions of stripe, checkerboard CO and LTO superstructure peaks. ({\bf b}) Scans along path A showing  magnetic and structural signals separated by polarization analysis. ({\bf c,d}) Magnetic and structural scattering measured in scans along paths B and C. In {\bf b} and {\bf d}, small spurious peaks from half wavelength $(\lambda/2)$ contamination in the neutron beam are indicated. The shaded peaks in {\bf d} indicate low-temperature orthorhombic superlattice reflections.}
\label{fig2}
\end{figure*}

Figure~\ref{fig2}a is a sketch of the $(h,h,l)$ plane in reciprocal space showing the main scattering features investigated here. The reciprocal lattice is defined with respect to the conventional tetragonal unit cell (space group I4/mmm, lattice parameters $a=0.386$\,nm, $c=1.26$\,nm) which describes the room temperature crystal structure of \LSCO. Diffraction scans along the paths marked A, B and C are presented in Figs.~\ref{fig2}b--d from measurements made at a temperature $T = 2$\,K. Scan A in Fig.~\ref{fig2}b shows two broad magnetic peaks centred close to $h = 1/3$ and 2/3, respectively. These derive from the diagonally-modulated AFM short-range order observed previously  \cite{boothroyd-nature-2011,cwik-prl-2009} which produces a four-fold pattern of magnetic diffraction peaks centred on the N\'{e}el AFM wavevectors ${\bf Q}_{\rm AF} = (h+1/2, k+1/2, l)$ with integer $h, k, l$. The magnetic peaks are displaced from ${\bf Q}_{\rm AF}$ by $\pm (\epsilon_{\rm so}/2, \epsilon_{\rm so}/2, 0)$ and $\pm (\epsilon_{\rm so}/2, -\epsilon_{\rm so}/2, 0)$, where $\epsilon_{\rm so} \approx x$ for doping levels between 0.25 and 0.5 (refs~\onlinecite{cwik-prl-2009,gaw-prb-2013}). From the fitted magnetic peak positions we find $\epsilon_{\rm so} = 0.37 \pm 0.01$, and from their widths (half width at half maximum) we determine a magnetic correlation length of $\xi_{\rm so} = 0.68\pm 0.06$\,nm in the $[110]$ direction, consistent with the previous unpolarized neutron diffraction study \cite{boothroyd-nature-2011}.

\begin{figure*}
\includegraphics[width=0.9\textwidth]{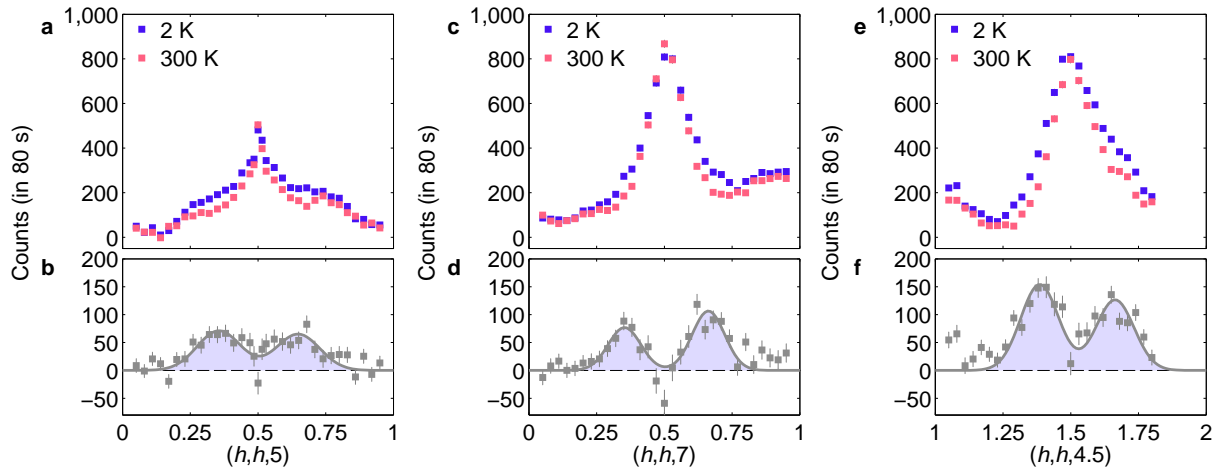}
\caption{\textbf{Charge order scattering from \LSCO\ }
Structural scattering intensities are compared at temperatures of 2\,K and 300\,K in scans along {\bf a},{\bf b} $(h,h,5)$, {\bf c},{\bf d} $(h,h,7)$ and {\bf e},{\bf f} $(h,h,4.5)$. The spurious peaks at $h \simeq 0.25$ \& 0.75 in {\bf a}, at $h \simeq 0.9$ in {\bf c}, and at $h \simeq 1.1$ \& 1.7 in {\bf e}, are aluminium powder peaks. The lower plots in each panel show the difference between the curves recorded at 2\,K and 300\,K. The shaded peaks are fits of two Gaussian peaks to this data.
}
\label{fig3}
\end{figure*}

The non-magnetic signal in scan A contains a peak at $h = 0.5$  with two small shoulders on its flanks at about the same positions as the magnetic peaks (Fig.~\ref{fig2}b). Figure~\ref{fig3}a compares the same scan at temperatures of 2\,K and 300\,K. The shoulders of scattering near $h=1/3$ and 2/3 are no longer present at 300\,K, whereas the amplitude of the peak at $h=0.5$ remains virtually unchanged. The difference between the 2\,K and 300\,K data sets, also shown in Fig.~\ref{fig3}b, reveals two peaks centred approximately on $h=1/3$ and 2/3. The presence of this additional structural scattering at low temperatures was confirmed in several other scans parallel to scan A, two of which are shown in Figs.~\ref{fig3}c-f. These reproduce the same key features as observed in Fig.~\ref{fig3}a,b: the central peak does not change in amplitude between 2\,K and 300\,K, but additional diffuse scattering peaks centred at $h \simeq 1/3$ and 2/3 develop below 300\,K.

This excess structural diffuse scattering at low temperature is consistent with the presence of charge stripes. In the stripe model the diagonal charge modulation gives rise to CO diffraction peaks which are displaced from the reciprocal lattice points by $\pm (\epsilon_{\rm co}, \epsilon_{\rm co}, 0)$, where $\epsilon_{\rm co} = \epsilon_{\rm so}$.  From our data we obtain $\epsilon_{\rm co} = 0.36 \pm 0.01$, in agreement with $\epsilon_{\rm so} = 0.37 \pm 0.01$ found earlier. The fact that $\epsilon_{\rm so}$ and $\epsilon_{\rm co}$ are greater than 1/3, the value for ideal period-3 stripes (Fig.~\ref{fig1}a), suggests either that the hole doping level of our crystal could be slightly in excess of 1/3, or that the presence of defects in an otherwise ideal period-3 stripe pattern causes an effective small incommensurability\cite{Savici}.

\begin{figure*}
\includegraphics[width=0.7\textwidth]{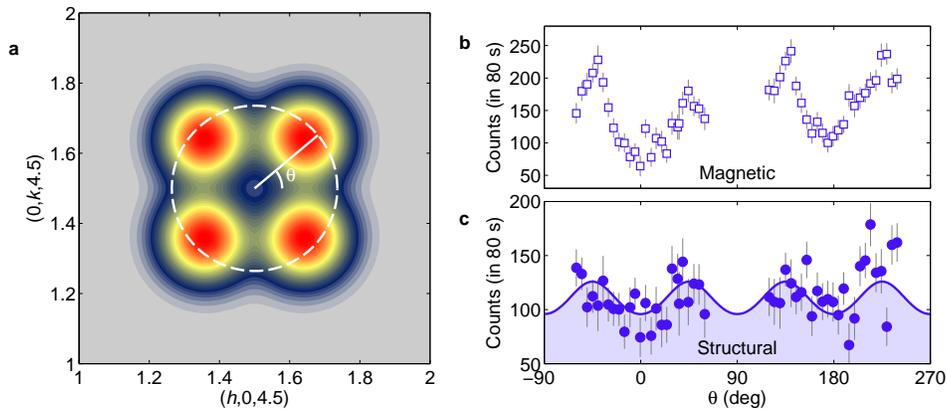}
\caption{\textbf{In-plane spin and charge scattering from stripes in \LSCO\ } ({\bf a}) Colour map of simulated charge stripe scattering intensity. ({\bf b}) magnetic and ({\bf c}) structural scattering around the circular path shown in {\bf a}. The intensity is the difference between measurements at 2\,K and 300\,K. The line and shading in {\bf c} are calculated from the fitted model shown in {\bf a}. }
\label{fig4}
\end{figure*}

Because the stripes can run along either diagonal of the square lattice with equal probability we also expect CO peaks at $\pm (\epsilon_{\rm co}, -\epsilon_{\rm co}, 0)$ relative to the reciprocal lattice points. This pair of peaks together with those at $\pm (\epsilon_{\rm co}, \epsilon_{\rm co}, 0)$ generate the characteristic fourfold pattern of CO peaks depicted in Fig.~\ref{fig4}a that accompanies the fourfold pattern of SO peaks. To demonstrate this tetragonal anisotropy we show in Figs.~\ref{fig4}b and c the magnetic and structural scattering intensities measured along two arcs of the circular path indicated on Fig.~\ref{fig4}a. We performed this type of scan by tilting the crystal to vary the wavevector component out of the $(h,h,l)$ scattering plane.  The magnetic and charge scans both display the same periodic variation in intensity around the circular path, with the maxima located at the positions expected for stripe order. In Fig.~\ref{fig4}a the charge peaks are depicted as isotropic (i.e.~circular), but diffuse scattering peaks from stripe order will in general display twofold symmetry as a result of different correlation lengths parallel and perpendicular to the stripes\cite{Savici}. The SO peaks display this expected twofold anisotropy as reported previously,\cite{boothroyd-nature-2011}. However, more complete data would be needed to resolve any such anisotropy in the correlation lengths for the charge order.

Having established the presence of the stripe phase in \LSCO\ we now develop a more complete picture of its complex ground state from the $l$ and temperature dependence of the diffuse scattering. Figure~\ref{fig2}c presents scans of the magnetic diffuse scattering at 2\,K along paths B and C. Scan B shows that the magnetic peaks are strongest at $l =$ odd integers in the $(h,h,l)$ plane. This is a consequence of the way the stripe order propagates along the $c$ axis (see Supplementary Note 1). The large peak widths in scan B imply that the magnetic correlations do not extend much beyond an adjacent layer. The magnetic scattering at ${\bf Q}_{\rm AF}$ (scan C) shows no periodic variation with $l$, consistent with a superposition of the four magnetic peaks which surround ${\bf Q}_{\rm AF}$, one pair of which has maxima at $l =$ odd integers and the other at $l =$ even integers\cite{boothroyd-nature-2011}.

The $l$ variation of the structural diffuse scattering at 2\,K shows a number of features of interest. Scan C (Fig.~\ref{fig2}d) contains relatively sharp peaks at $l = 2, 4, 6$ and 8 with some diffuse scattering in between which is strongest near $l=5$ and 7. The sharp peaks at $l =$ even integers  are superstructure reflections from the low temperature orthorhombic (LTO) structural distortion which occurs in \LSCOx\ for $x \lesssim 0.4$ (refs~\onlinecite{yamada-prb-1989,cwik-prl-2009}). The intervening diffuse scattering has a similar variation with $l$ to that observed in the equivalent scan from the checkerboard CO phase in La$_{3/2}$Sr$_{1/2}$CoO$_4$ (refs~\onlinecite{zaliznyak-prl-2000,zaliznyak-prb-2001,zaliznyak-jap-2004}). This confirms the presence of checkerboard CO in \LSCO, as reported previously \cite{drees-nature-2014}, and accounts for the peaks at $h=0.5$ and 1.5 in the scans shown in Fig.~\ref{fig2}b and Fig.~\ref{fig3}. The observed $l$-dependence of this checkerboard CO scattering has been reproduced by a model for the pattern of subtle displacements of the in-plane and apical oxygen ions that surround the Co$^{2+}$ and Co$^{3+}$ sites \cite{zaliznyak-prl-2000,zaliznyak-prb-2001,zaliznyak-jap-2004}.

Scan B (Fig.~\ref{fig2}d), in which roughly half the intensity is from the stripe CO and half from the tail of the checkerboard CO peak, exhibits a similar $l$-dependence to that from the checkerboard CO in scan C. Although the stripe CO and checkerboard CO peaks occur at different in-plane wave vectors, their $l$-dependence is expected to be similar because in both cases it originates from much the same local displacements of the oxygen ions in the CoO$_6$ octahedra.  The structure factor of this local distortion mode is very small for $l \leq 3$ (refs~\onlinecite{zaliznyak-prl-2000,zaliznyak-prb-2001,zaliznyak-jap-2004}), which may explain why previous attempts to detect stripe CO in \LSCO\ from measurements at $l \leq 3$ have been unsuccessful.

\subsection*{Temperature evolution of stripe spin and charge order}

\begin{figure}
\includegraphics[width=\columnwidth]{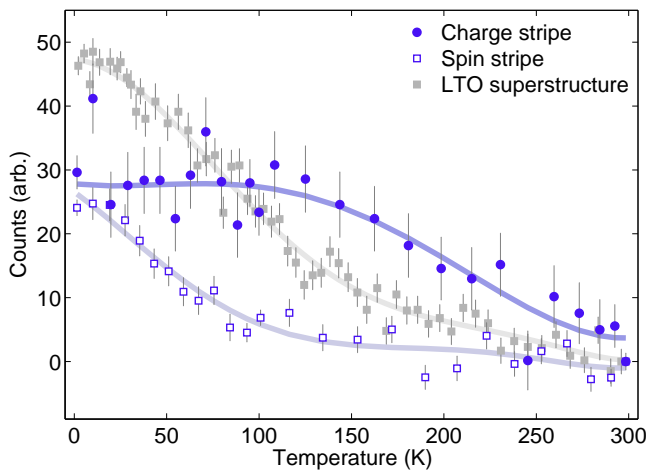}
\caption{\textbf{Temperature dependence of diffraction peaks}
The LTO superstructure data are from measurements at $(0.5,0.5,4)$, and the spin and charge stripe data are from measurements at $(1.38,1.38,4.5)$. In each case, a measurement made at 300\,K has been subtracted from the remaining data. The lines are guides for the eye. }
\label{fig5}
\end{figure}

Figure~\ref{fig5} shows the temperature dependence of the stripe and LTO superstructure peaks found in \LSCO. Charge stripe correlations develop continuously on cooling from room temperature down to 100\,K, below which the signal remains constant. This temperature variation is consistent with the observation of charge freezing at $\sim 100$\,K in magnetic resonance experiments\cite{lancaster-prb-2014,williams-arxiv-2015}. The associated spin stripe scattering begins to increase below about 100\,K. The separation $\epsilon_{\rm so}$ of the magnetic peaks changes only slightly between 2\,K and $\sim$90\,K (see Supplementary Note 2), consistent with the notion that the magnetic peaks are associated with the charge stripe order which is fully developed at 100\,K. A similar behaviour was observed in La$_{5/3}$Sr$_{1/3}$NiO$_4$ (ref.~\onlinecite{lee-prl-1997}). The LTO superstructure appears gradually below room temperature but develops most strongly below about 150\,K, in accord with the published phase diagram of \LSCOx\  (ref.~\onlinecite{cwik-prl-2009}) and previous results for $x = 0.33$ (ref.~\onlinecite{drees-nature-2014}).

As already seen in Fig.~\ref{fig3}, the checkerboard CO does not change below room temperature in \LSCO. This is consistent with the very robust nature of the checkerboard phase, which persists up to about $825$\,K in half-doped La$_{3/2}$Sr$_{1/2}$CoO$_4$ (ref.~\onlinecite{zaliznyak-prb-2001}).  In La$_{3/2}$Sr$_{1/2}$CoO$_4$, AFM ordering of spins on the Co$^{2+}$ sites (Fig.~\ref{fig1}b) sets in gradually at around 40\,K and gives rise to strong diffraction peaks close to $(0.25, 0.25, l)$ and $(0.75, 0.75, l)$ with $l$ odd\cite{zaliznyak-prb-2001,helme-prb-2009}. We found no evidence for such AFM order associated with the checkerboard CO phase in \LSCO.

\section*{Discussion}
Our measurements indicate that the ground state of \LSCO\ is phase-separated into two dominant components with different local hole concentrations: (1) spin-and-charge stripe order ($x \simeq 1/3$), and (2) checkerboard charge order ($x \simeq 1/2$). The typical size of the correlated regions is $\lesssim 1$\,nm. At low temperature the different orders appear static on the timescale of neutron diffraction ($\sim 10^{-12}$\,s) and probably also magnetic resonance\cite{lancaster-prb-2014,williams-arxiv-2015} ($\sim 10^{-6}$\,s), but given the short correlation lengths they must in reality be quasi-static, i.e.~not static but fluctuating at a rate that is slower than the experimental time window. To maintain charge neutrality there must also be regions with local doping $x < 1/3$, but these regions must be small as we could not detect any distinct diffraction signature of them. In particular, we did not observe any magnetic peaks at ${\bf Q}_{\rm AF}$ characteristic of La$_2$CoO$_4$-type N\'{e}el AFM order.

The results obtained here are significant for understanding the ground state and magnetic dynamics of \LSCOx, and could also have implications for the copper oxide high temperature superconductors. The magnetic spectrum of \LSCOx\ for $x=0.25-0.4$ is notable in that its overall structure, i.e.~the distribution of intensity as a function of energy and in-plane wavevector observed in neutron scattering measurements, has an hourglass shape \cite{boothroyd-nature-2011,gaw-prb-2013,drees-nature-2013}. The spectrum emerges from the four incommensurate magnetic diffraction peaks which surround ${\bf Q}_{\rm AF}$. With increasing energy the four peaks first disperse inwards until they merge at ${\bf Q}_{\rm AF}$ then disperse outwards again. Above the waist of the hourglass the spectrum also has a four-fold pattern but now rotated by 45$^{\circ}$ with respect to the pattern below the waist.

The discovery of quasi-static charge stripes consolidates the stripe scenario proposed to explain the hourglass spectrum in \LSCO\ (ref.~\onlinecite{boothroyd-nature-2011}). An hourglass spectrum arises naturally from a stripe-ordered ground state in which magnetic correlations are stronger along the stripes than across them\cite{yao-prl-2006,andrade-prl-2012}. In such a case, the lower and upper parts of the spectrum are from the inter- and intra-stripe parts of the magnon dispersion, respectively, and the waist is a saddle point formed from the maximum of the inter-stripe dispersion and the minimum of the intra-stripe dispersion. The 45$^{\circ}$ rotation of the intensity maxima above the waist relative to the fourfold pattern below the waist is explained by the superposition of quasi-one-dimensional dispersion surfaces from orthogonal stripe domains\cite{boothroyd-nature-2011}.

Recently, a stripe-free model of \LSCOx\ has been proposed that also has an hourglass spectrum \cite{drees-nature-2013,drees-nature-2014}. The model is based on a phase-separated ground state comprising nanoscopic $x=0.5$ regions with checkerboard CO coexisting with undoped ($x=0$) regions with N\'{e}el AFM order \cite{drees-nature-2013}. In the this picture, the part of the hourglass spectrum below the waist derives from magnetic correlations associated with the checkerboard CO regions while the part above waist comes from  N\'{e}el AFM correlations in the the undoped regions.


Let us consider the stripe-free model in the light of the present results. First, in our sample of \LSCO\  we found $\epsilon_{\rm so} \simeq \epsilon_{\rm co} = 0.36 \pm 0.01$, which is significantly different from the value $\epsilon_{\rm so} = \epsilon_{\rm co} = 0.5$ characteristic of the AFM order associated with the ideal checkerboard CO (Fig.~\ref{fig1}b). In our view, this is clear evidence that the strong magnetic peaks from which the low energy part of the hourglass spectrum emerges are associated with the stripe component of the ground state and not the checkerboard regions. Second, this evidence is reinforced by the fact that the magnetic diffraction peaks in \LSCO\ persist up to approximately 100\,K (Fig.~\ref{fig5} and Supplementary Fig.~4), well above the corresponding magnetic ordering temperature $\simeq 40$\,K of the checkerboard CO phase observed of La$_{3/2}$Sr$_{1/2}$CoO$_4$ (refs~\onlinecite{zaliznyak-prl-2000,helme-prb-2009}). Due to disorder effects, the magnetic correlations are expected to be weaker in nano-sized patches of checkerboard CO than in bulk La$_{3/2}$Sr$_{1/2}$CoO$_4$, and the magnetic dynamics of these spins could be sufficiently slow and diffusive that they do not influence the observed hourglass spectrum, consistent with the idea that the observed spectrum is dominated by that from the stripe phase. Third, the absence of La$_2$CoO$_4$-type AFM order means there can be no gap in the spin fluctuation spectrum associated with spin correlations in any undoped regions, since a gap requires broken rotational symmetry. This is inconsistent with the stripe-free model, which requires the fully-gapped La$_2$CoO$_4$-type magnon spectrum in order to produce the waist and upper part of the hourglass. In addition, the magnon spectrum of N\'{e}el-ordered La$_2$CoO$_4$ is isotropic until the spectrum approaches the zone boundary at high energies \cite{babkevich-prb-2010}, and therefore cannot reproduce the observed tetragonal anisotropy and 45$^{\circ}$ rotation of the intensity maxima relative to the fourfold pattern below the waist of the hourglass \cite{boothroyd-nature-2011,gaw-prb-2013}.

Although the ground state of \LSCO\ is more complex than originally assumed, we conclude that the weight of evidence favours the stripe explanation for the hourglass spectrum.   Many hole-doped copper oxide superconductors exhibit an hourglass spectrum with the same general structure as that of \LSCO\ (ref.~\onlinecite{fujita-jpsj-2012}), and some underdoped cuprates also have similar (quasi-)static spin and charge stripe order\cite{tranquada-nature-1995,kivelson-rmp-2003}. The magnetic spectrum is important for understanding the physics of cuprates because it offers a window onto the nature of the correlated magnetic ground state which supports superconductivity. The present results provide an experimental basis for theories that assume a ground state with static or slowly fluctuating stripes in order to explain the hourglass spectrum in cuprates.

\section{Methods}
\subsection{Experimental details}

The single-crystal sample of \LSCO\ was grown by the floating-zone method in an image furnace and is part of the larger of the two crystals measured previously \cite{boothroyd-nature-2011}. Polarized neutron diffraction measurements were performed with the longitudinal polarization setup of the IN20 triple-axis spectrometer at the Institut Laue--Langevin. Approximately 5\,g of the crystal was mounted in a helium cryostat and aligned with $(h,h,l)$ as the horizontal scattering scattering plane. Incident and scattered neutrons of energy 14.7\,meV were selected by a Heusler monochromator and a Heusler analyser. Two pyrolytic graphite (PG) filters were placed before the sample to suppress second- and third-order harmonic contamination in the incident beam. A magnetic guide field at the sample position ensured that the direction of the neutron spin polarization ($\bf P$) was aligned with the neutron scattering vector ($\bf Q$), and a spin flipper was placed after the sample. Neutrons whose spin direction reversed (spin-flip scattering) or did not reverse (non-spin-flip scattering) on scattering were recorded separately. Measurements were recorded for 80\,s per point per polarization channel, but scans shown in Figs.~\ref{fig3}--\ref{fig4} were repeated several times and averaged. Error bars on the data in Figs.~\ref{fig2}--\ref{fig5} are standard deviations obtained from neutron counts.

\subsection{Corrections for non-ideal neutron polarization}

In principle, polarization analysis can be used to completely separate magnetic of electronic origin from structural scattering because when the neutron polarization $\bf P$ is parallel to the scattering vector $\bf Q$ the spin of the neutron is always flipped in a magnetic interaction and is always unchanged in a coherent nuclear interaction\cite{moon-riste-koehler-1969}. In reality, inaccuracies due to imperfect neutron polarization and flipping efficiency  cause some leakage of the magnetic scattering into the non-spin-flip channel and \textit{vice versa}. The quality of the polarization setup is represented by the flipping ratio $R$, which in our experiment was measured on a nuclear Bragg peak: $R = (I_{\rm Bragg}^{\rm NSF}-B^{\rm NSF})/(I_{\rm Bragg}^{\rm SF}-B^{\rm SF})$. Here, $I^{\rm NSF}$ and $I^{\rm SF}$ are the raw counts in the non-spin-flip (NSF) and spin-flip (SF) channels, and $B^{\rm NSF}$ and $B^{\rm SF}$ are corresponding backgrounds (incoherent scattering and instrumental backgrounds). The observed value was $R = 25.7$, which corresponds to 92.5\% polarisation.

Supplementary Fig. 1 provides an illustration of the separation of magnetic and nuclear scattering before any corrections were made. The plot shows the raw intensities $I^{\rm NSF}$ and $I^{\rm SF}$ measured  along the line $(h,h,3)$ in reciprocal space. The structural diffuse scattering from charge order and the LTO distortion is very small in this particular scan, and the near absence of any structure in $I^{\rm NSF}$  shows that there is very little leakage of the strong magnetic peaks into the NSF channel. The peaks fitted to $I^{\rm NSF}$ and $I^{\rm SF}$ are consistent with $R = 25.7$, which demonstrates that the good polarisation observed in Bragg diffraction also extends to diffuse scattering.

As there inevitably is a small leakage between channels, which is almost entirely due to beam polarization, we applied standard corrections to the data given by
\begin{eqnarray}
\left( \hspace{-4pt} \begin{array}{l} N\\[3pt] M\end{array} \hspace{-4pt} \right)_{\hspace{-2pt}\rm corr}  =  \frac{1}{R-1}
 \left( \hspace{-4pt}\begin{array}{c c}
                  R & -1 \\[2pt]
                  -1 & R \\[2pt]
                  \end{array}\right)
                  \left( \hspace{-4pt} \begin{array}{l} I^{\rm NSF} - B^{\rm NSF}\\[3pt] I^{\rm SF} - B^{\rm SF}\end{array} \hspace{-4pt} \right)_{\hspace{-2pt}\rm meas},\nonumber
\label{eq:P-corrections}
\end{eqnarray}
where $N$ and $M$ are the corrected nuclear and magnetic intensities. $M$ contains the total intensity from  magnetic components perpendicular to $\bf Q$.

Supplementary Fig. 2 illustrates the effect these corrections have on the structural diffuse scattering studied in this work. The upper panels (a, d and g) show uncorrected data. The middle panels (b, e, h) show data corrected with $R=25.7$, and the lower panels (c, f, i) are corrected with $R=15$, a value well below that observed in the experiment. In each case, data recorded at 2\,K and 300\,K are shown in the first two columns, and the difference between 2\,K and 300\,K is plotted in the third column for the NSF/$N$ channel. The effect of the polarization corrections is seen to be negligible. This is important because at low temperature the structural diffuse scattering peaks due to the stripe charge order are almost coincident with the magnetic diffraction peaks. The results in Supplementary Fig.~2 demonstrate that the corrections are not very sensitive to the value of $R$, and so one can be confident that the stripe CO signal obtained in this work is not the result of feed-through from the magnetic scattering channel.

\section{Data availability}
The neutron diffraction data that support the findings of this study are available from the Institut Laue--Langevin with the identifiers [https://dx.doi.org/10.5291/ILL-DATA.5-53-247 and https://dx.doi.org/10.5291/ILL-DATA.DIR-132]\cite{in20-1,in20-2}.

\section{Acknowledgements}
This work was supported by a European Research Council CONQUEST grant and by the Engineering \& Physical Sciences Research Council of the United Kingdom (grant no. EP/J012912/1).

\section{Competing financial interests}
The authors declare no competing financial interests.

\section{Author contributions}
D.P. prepared and characterized the single-crystal samples. A.T.B., P.B., P.G.F. and M.E. performed the neutron scattering experiments. P.B. performed the data analysis, and P.B. and A.T.B. wrote the manuscript with input from the other co-authors.

\newpage

\begin{figure*}[hp!]
\begin{center}
\includegraphics[width=0.45\textwidth]{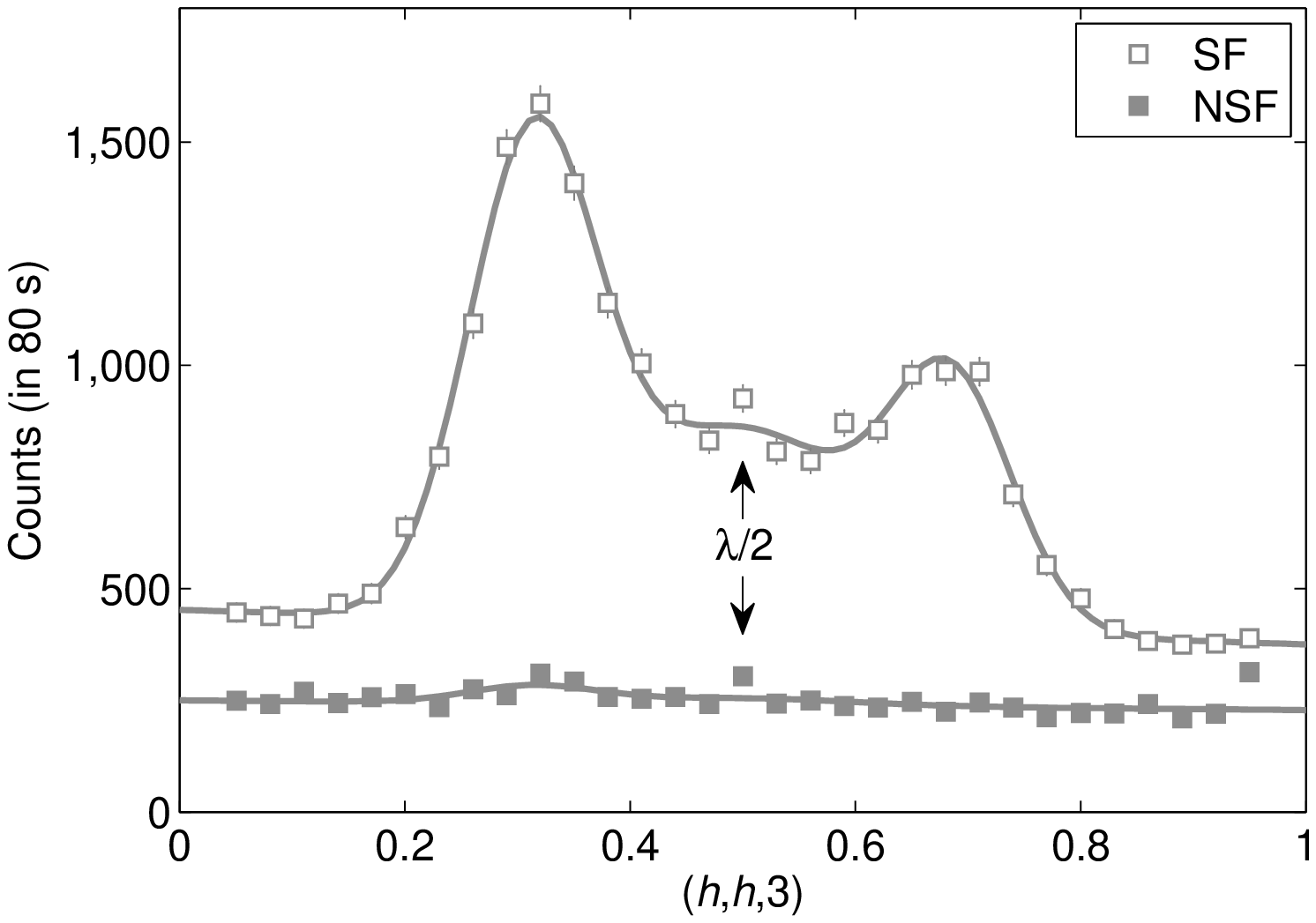}
\end{center}
\justify{\label{figS1}
\textbf{Supplementary Figure 1: Separation of magnetic and nuclear scattering. }
Raw polarized neutron diffraction intensities for \LSCO\ measured at a temperature of 2\,K along the line $(h,h,3)$ in reciprocal space. The two peaks in the SF channel are from the stripe magnetic order.  There is very little leakage between the two polarization channels, consistent with the assumed flipping ratio $R=25.7$. The small peak at $h = 0.5$ in both channels is due to second order diffraction ($\lambda/2$) from the $(1,1,6)$ Bragg peak. This and other peaks originating from $\lambda/2$ contamination are temperature-independent and so cancel out when scans at different temperatures are subtracted, as was done to reveal the stripe CO scattering. Error bars on the data are standard deviations obtained from neutron counts.}
\end{figure*}

\begin{figure*}[hp!]
\begin{center}
\includegraphics[width=0.85\textwidth,trim= 0pt 0pt 0pt 0pt, clip]{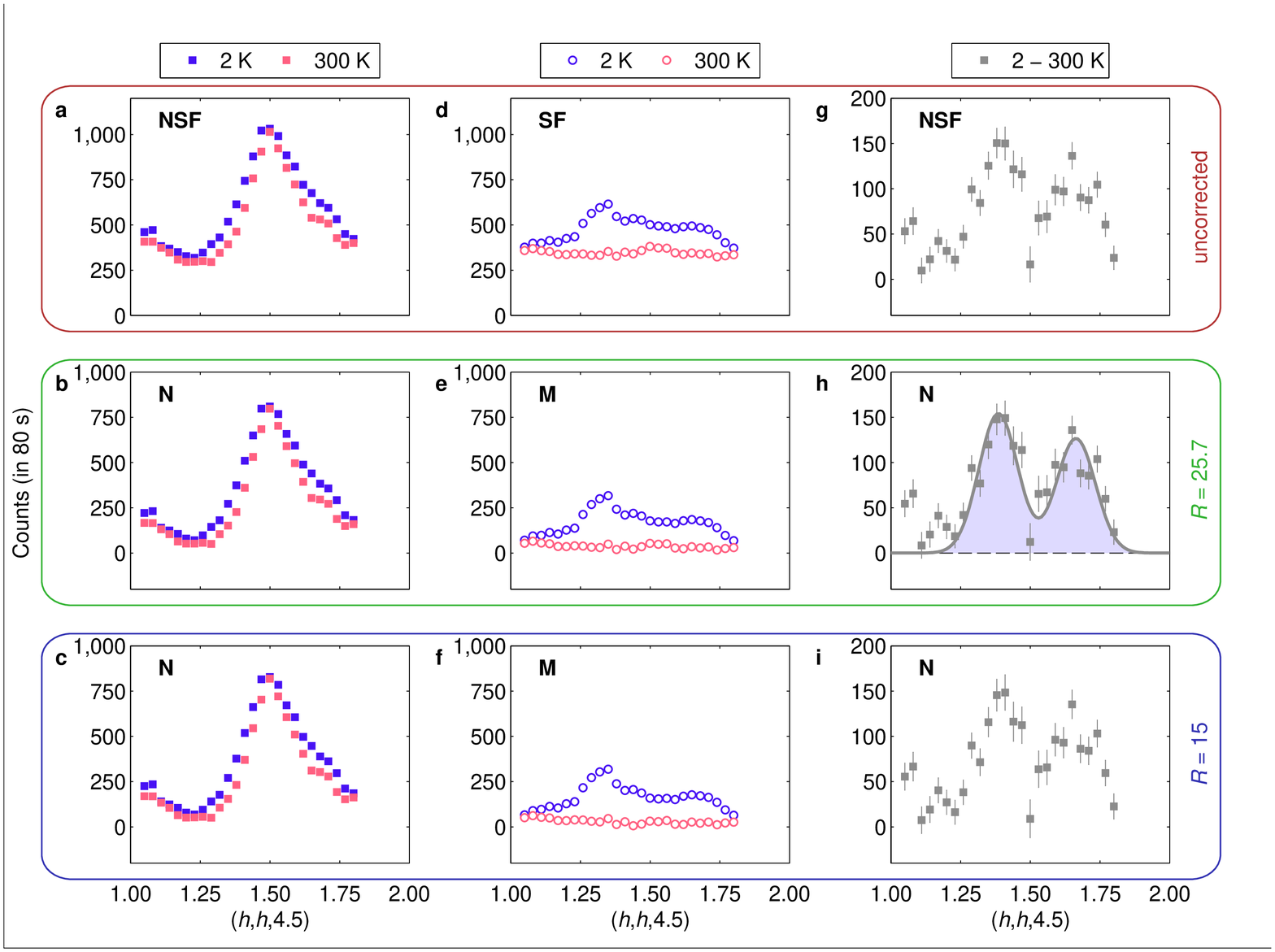}
\end{center}
\justify{\label{figS2}
\textbf{Supplementary Figure 2: Effect of polarisation corrections.}
The upper three panels show raw, uncorrected data. The middle and lower three panels show data corrected with $R=25.7$ and $R=15$, respectively. Panels {\bf a} -- {\bf c} contain the raw NSF scattering and corrected structural diffuse scattering ($N$) due to checkerboard and stripe charge order, and panels {\bf d} -- {\bf f} contain the raw SF scattering and corrected magnetic diffuse scattering ($M$). Panels {\bf g} -- {\bf i} show the raw and corrected NSF signal after subtraction of the 300\,K  data from the 2\,K data to isolate the structural diffuse scattering due to stripe charge order. Error bars on the data are standard deviations obtained from neutron counts.}
\end{figure*}

\clearpage

\begin{figure}
\begin{center}
\includegraphics[width=0.9\columnwidth,trim= 0pt 0pt 0pt 0pt, clip]{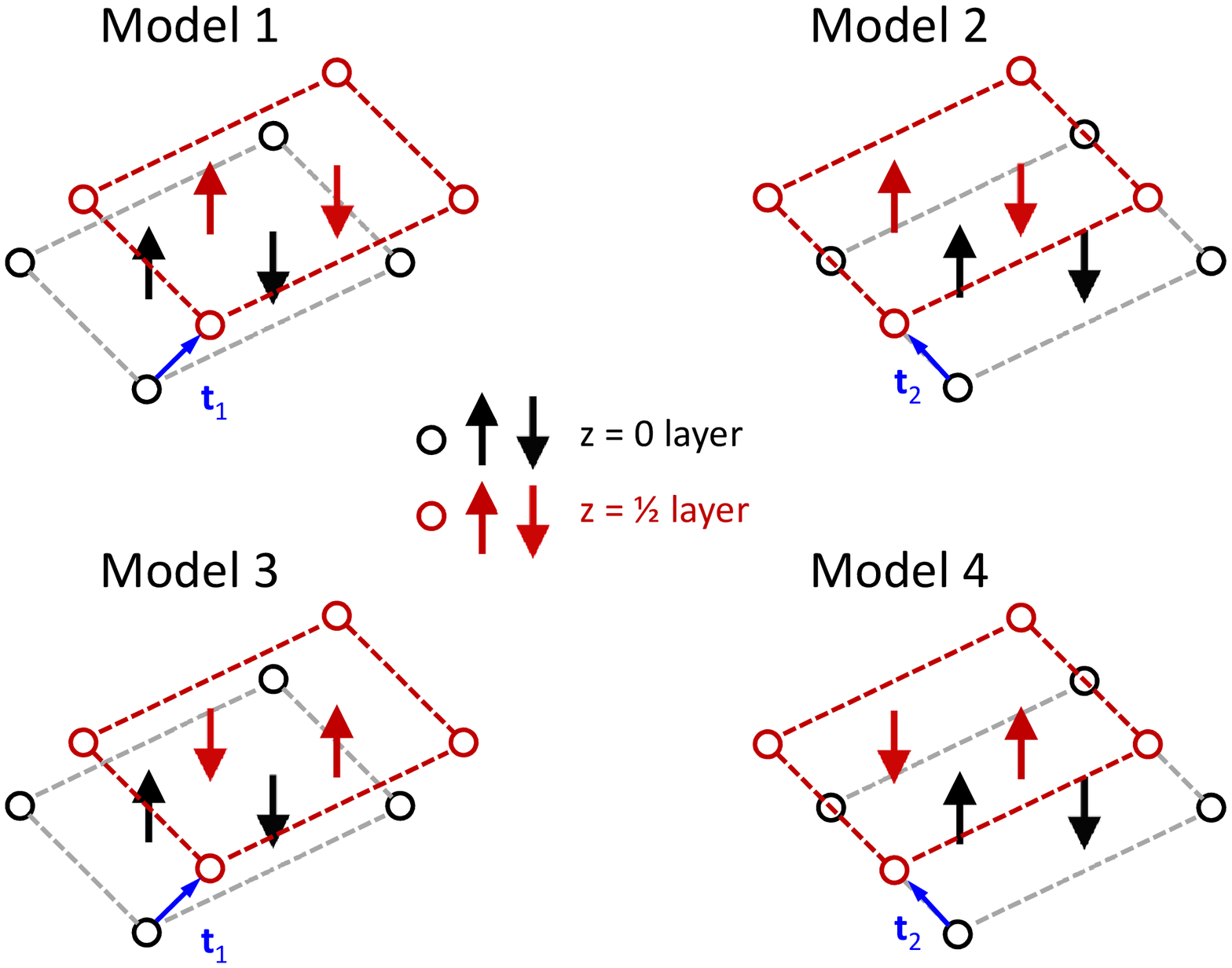}
\end{center}
\justify{\label{figS4}
\textbf{Supplementary Figure 3: Models for the spin and charge order on adjacent layers in \LSCO.} The open circles represent Co$^{3+}$ holes, and the arrows indicate ordered spins on Co$^{2+}$. Black and red symbols are for $z=0$ and $z = 1/2$ layers, respectively. ${\bf t}_1$ and ${\bf t}_2$ are the two different stacking vectors. The parallelograms are unit cells for the spin and charge order within each CoO$_2$ layer.}
\end{figure}

\begin{table}[hp!]
\justify{\label{CO-SF}
\textbf{Supplementary Table 1.}
Values of the factor $1 + \exp ({\rm i}{\bf Q} \cdot {\bf t})$ for charge order (CO) diffraction peaks for each of the four ideal period-3 stripe structure shown in Supplementary Fig.~3. Values are given for two of the CO peaks surrounding $(0.5,0.5,l)$, one from each stripe domain.}
\begin{center}
\begin{tabular}{c c c c c c}
\hline
\hline
\textrm{CO} & stacking & \multicolumn{2}{c}{$(1/3,1/3,l)$} &  \multicolumn{2}{c}{$(2/3,1/3,l)$}  \\
 model& ${\bf t}$ & $l$ even & $l$ odd & $l$ even & $l$ odd \\
\hline
& & & & &\\[1pt]
1 \& 3 & ${\bf t}_1$ & 1 & 3 & 3 & 1 \\[3pt]
2 \& 4 & ${\bf t}_2$ & 4 & 0 & 0 & 4\\[2pt]
\hline
\hline
\end{tabular}
\end{center}
\end{table}
\begin{table}[hp!]
\justify{\label{SO-SF}
\textbf{Supplementary Table 2.}
Values of the factor $1 + T\exp ({\rm i}{\bf Q} \cdot {\bf t})$ for spin order (SO) diffraction peaks for each of the four ideal period-3 stripe structures shown in Supplementary Fig.~3. Values are given for two of the SO peaks surrounding $(0.5,0.5,l)$, one from each stripe domain.}
\begin{center}
\begin{tabular}{cccccc}
\hline
\hline
\textrm{SO} & stacking & \multicolumn{2}{c}{$(1/3,1/3,l)$}  & \multicolumn{2}{c}{$(2/3,1/3,l)$}  \\
 model & $({\bf t}, T)$ & $l$ even & $l$ odd & $l$ even & $l$ odd \\
\hline
1 & $({\bf t}_1, +1)$ & 1 & 3 & 3 & 1 \\[3pt]
2  & $({\bf t}_2, +1)$ & 4 & 0 & 0 & 4\\[3pt]
3 & $({\bf t}_1, -1)$ & 3 & 1 & 1 & 3 \\[3pt]
4 & $({\bf t}_2, -1)$ & 0 & 4 & 4 & 0\\[2pt]
\hline
\hline
\end{tabular}
\end{center}
\end{table}
\newpage

\section*{Supplementary Note 1: \\Stacking of spin and charge stripe order along c axis}

In this section we analyze the stacking of the incommensurate CO and SO based on the scans presented in Fig.~2c,d in the main article, which show that spin and charge correlations extend over a short distance $\sim c$ perpendicular to the CoO$_2$ layers. Since both the charge and magnetic peaks are centred at integer $l$ values the majority stacking of the (quasi-)order has the same $c$-axis period as the lattice. However, there are several different ways in which the order can be related on adjacent layers related by the body-centering translation.

We consider ideal site-centred period-3 stripes and assume that the magnetic structure is collinear. This means that the charge superstructure on one layer ($z=0$) can be related to that on an adjacent layer ($z=1/2$) by a translation $\bf t$.  For the spin order, the translation is followed by the time reversal operation $T =\pm 1$, where $T = +1$ means no change in the spins, and $T = -1$ means reversal of all the spins. The structure factors for the spin and charge diffraction peaks then contain a factor $1 + T\exp ({\rm i}{\bf Q} \cdot {\bf t})$, where $T = +1$ for charge order and $T= \pm1$ for spin order. In the exponential, ${\bf Q} \cdot {\bf t} = (Q_xt_x+Q_yt_y+Q_zt_z)$, where $t_x$, $t_y$, $t_z$ are the components of $\bf t$ written as fractional coordinates along the crystallographic $a$, $b$, and $c$ axes, and $Q_x$, $Q_y$, $Q_z$ are the components of {\bf Q} in reciprocal lattice units.

There are two distinct stackings of the charge order, which may be described by stacking vectors ${\bf t}_1 = (0.5,0.5,0.5)$ and ${\bf t}_2=(-0.5,0.5,0.5)$. For each of these, the magnetic stacking can have $T = +1$ or $T = -1$. This gives four distinct superstructures for the combined spin and charge order with the periodicity of the lattice in the $c$ direction. The unit cells of each of these are shown in Supplementary Fig.~3 for a stripe domain in which the stripes run parallel to the $[1,-1]$ direction on the square lattice. Rotation of the stripe pattern by 90$^{\circ}$ gives an equivalent domain in which the stripes are parallel to the $[1,1]$ direction.

For ideal period-3 stripes the charge order (CO) peaks are located at ${\bf Q}_{\rm co} = {\bf G} \pm {\bf q}_1$ and ${\bf G} \pm {\bf q}_2$, where ${\bf G} = (h,k,l)$ is a reciprocal lattice vector, and the spin order (SO) diffraction peaks are located at ${\bf Q}_{\rm so} = {\bf Q}_{\rm AF} \pm {\bf q}_1/2$ and ${\bf Q}_{\rm AF} \pm {\bf q}_2/2$, where ${\bf Q}_{\rm AF} = (h+1/2, k+1/2, l)$ is an antiferromagnetic wavevector. ${\bf q}_1=(1/3,1/3,0)$ and ${\bf q}_2=(1/3,-1/3,0)$ are modulation wavevectors for the two equivalent orthogonal stripe domains. In the ${\bf q}_1$ domain the stripes run parallel to the $[1,-1]$ direction on the square lattice, and in the ${\bf q}_2$ domain the stripes are parallel to the $[1,1]$ direction.

Supplementary Tables~\ref{CO-SF} and \ref{SO-SF} list the values of the $1 + T\exp ({\rm i}{\bf Q} \cdot {\bf t})$ factor for the quartets of CO and SO satellites surrounding $(0.5,0.5,l)$, $l$ even and odd. The satellite peaks $(1/3,1/3,l)$ and $(2/3,1/3,l)$ are representative of the two domains. The data recorded along scan B presented in Fig.~2c in the main article show that for the $(1/3,1/3,l)$-type peaks the SO diffuse scattering is strongest for odd $l$ integers. From Supplementary Table~\ref{SO-SF} one can see that this observation favours the combined spin and charge stripe stacking described by either model 1 or model 4. Scan B in Fig.~2d indicates that the CO diffuse scattering has maxima close to odd $l$ and minima at even $l$, which from Supplementary Table~\ref{CO-SF} favours models 1 and 3. Taken together, therefore, the evidence supports model 1. However, this conclusion should be considered tentative as the peaks are very broad in $l$ and it is possible that the observed $l$-dependence comes entirely from the form factor of the CoO$_6$ distortion pattern, and not from any inter-layer correlations. A direct way to check this would be to measure the $l$-dependence along $(2/3, 1/3, l)$. If inter-layer correlations are significant then the non-magnetic scattering along $(2/3, 1/3, l)$ should be different from that along $(1/3, 1/3, l)$ in accord with the structure factors given in Supplementary Table~\ref{SO-SF}. Temperature difference data to fully isolate the $l$-dependence of the stripe CO scattering from the tails of the checkerboard CO peaks would also be valuable.

\begin{figure}
\includegraphics[width=0.65\columnwidth]{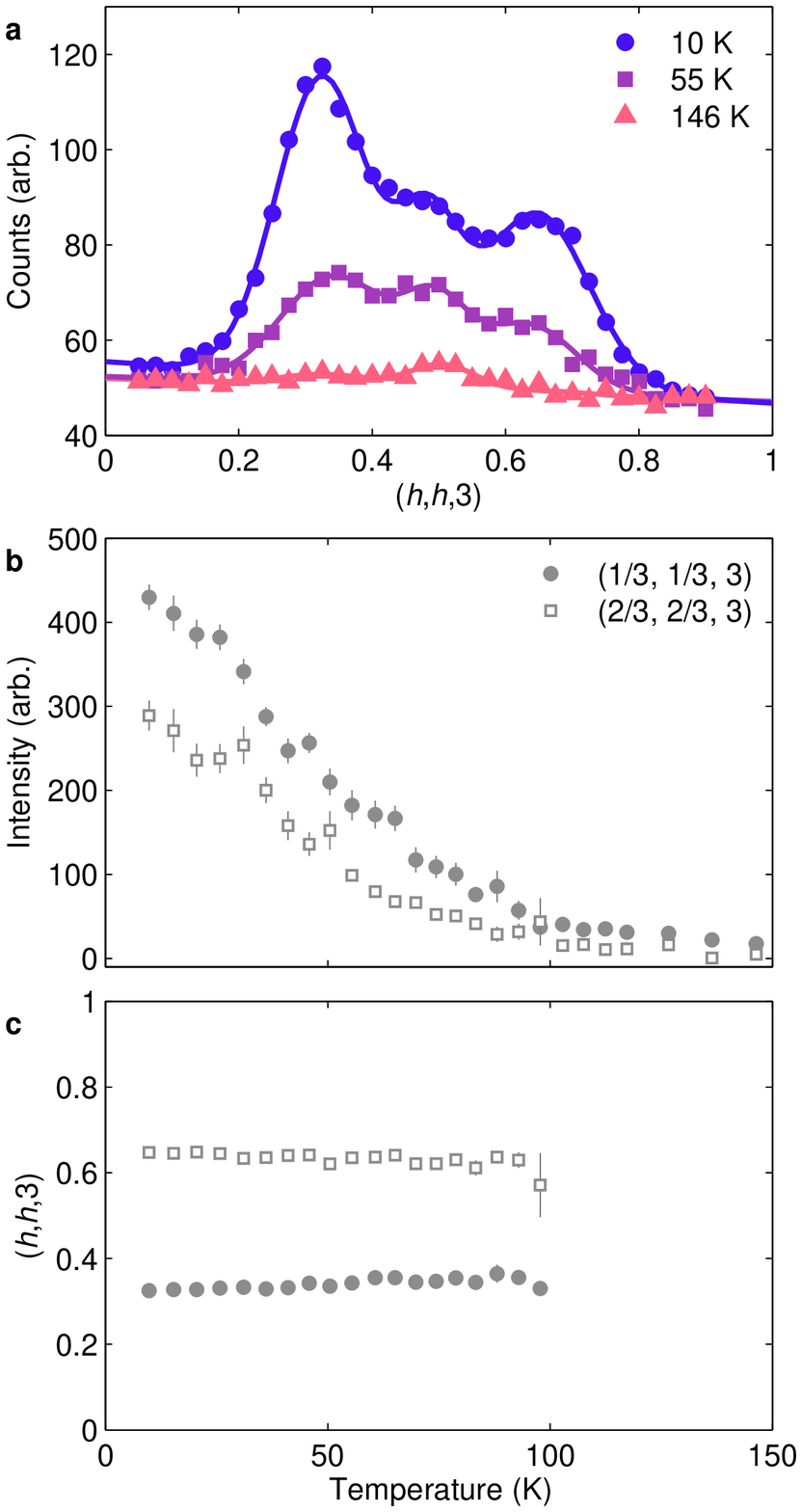}
\justify{\label{figS3}
\textbf{Supplementary Figure 4: Temperature dependence of magnetic order in \LSCO.}
({\bf a}) Unpolarised neutron diffraction measurements at 10\,K, 55\,K and 146\,K along the line $(h,h,3)$ in reciprocal space. ({\bf b}) Temperature dependence of the intensity of the fitted magnetic peaks. ({\bf c}) Temperature dependence of the magnetic peak positions.  Above 100\,K the magnetic signal is too weak to fit the peak centres. To estimate the magnetic intensity above 100\,K the peaks centres and widths were fixed, and only their amplitudes varied. }
\end{figure}

\newpage
\section*{Supplementary Note 2: \\Temperature dependence of magnetic order}

The temperature dependence of the magnetic diffraction peaks was studied by unpolarized neutron diffraction on the IN8 triple-axis spectrometer at the Institut Laue--Langevin. Incident and scattered neutrons with energy 14.7\,meV were selected by Bragg reflection from a Si $(111)$ monochromator and a pyrolytic graphite $(002)$ analyzer, respectively. The resolution will be slightly different from that achieved on IN20 owing to the different monochromator and analyzer materials and the fact that both were doubly focussed on IN8 whereas the IN20 analyser was only focussed horizontally.

Scans along $(h,h,3)$ were performed at a series of temperatures between 10\,K and 146\,K. Examples are shown in Supplementary Fig.~4a. From the polarized neutron measurements (Supplementary Fig.~1) there is very little structural diffuse scattering in this scan, so the temperature dependence of the scattering can be assumed to be from the magnetic order.  The temperature dependence of the peak intensities and positions are shown in Supplementary Figs.~3b and 3c. These were obtained by fitting a lineshape comprising three Gaussians to the data. By 100\,K the magnetic diffraction intensity has dropped to below 10\% of its value at the lowest temperature but the peak positions have shifted only slightly, towards the antiferromagnetic wavevector at the centre of the scan. This indicates that the incommensurate magnetic order is coupled to the incommensurate charge order, which is almost temperature independent below 100\,K (see Fig.~5 in the main article).

\end{document}